\RequirePackage{fix-cm}
\documentclass{article}       
\usepackage{graphicx, authblk}
\usepackage{mathptmx}      
\usepackage{amssymb}
\usepackage{amsmath}
\usepackage{subcaption}
\usepackage[sort,round]{natbib}
\usepackage{relsize}
\usepackage{lineno}
\usepackage{tikz}
\usepackage{tkz-fct}
\usetikzlibrary{math}
\usepackage{pgfplots}
\pgfplotsset{compat=1.15}
\usepackage{hyperref}

\newcommand{\possessivecite}[1]{\citeauthor{#1}'s (\citeyear{#1})}

\begin{document}

\title{Why males compete rather than care, with an application to supplying collective goods}

\author[1]{Sara L. Loo\thanks{S. Loo and D. Rose contributed comparably.}}  
\author[2]{Danya Rose}  
\author[3]{Michael Weight}  
\author[3]{Kristen Hawkes}  
\author[2]{Peter S. Kim}  


\affil[1]{School of Biotechnology and Biomedical Sciences, University of New South Wales, Sydney, NSW 2052, Australia}
\affil[2]{School of Mathematics and Statistics, University of Sydney, Sydney, NSW 2006, Australia}
\affil[3]{Department of Anthropology, University of Utah, Salt Lake City, UT 84112, USA }

\date{}

\maketitle

\begin{abstract}
The question of why males invest more into competition than offspring care is an age-old problem in evolutionary biology. On one hand, paternal care could increase the fraction of offspring surviving to maturity. On the other hand, competition could increase the likelihood of more paternities and thus the relative number of offspring produced. While drivers of these behaviours are often intertwined with a wide range of other constraints, here we present a simple dynamic model to investigate the benefits of these two alternative fitness-enhancing pathways. Using this framework we evaluate the sensitivity of equilibrium dynamics to changes in payoffs for male allocation to mating versus parenting. Even with strong effects of care on offspring survivorship, small competitive benefits can outweigh benefits from care. We consider an application of the model that includes men's competition for hunting reputations where big game supplies a benefit to all, and find a frequency-dependent parameter region within which, depending on initial population proportions, either strategy may outperform the other. Results demonstrate that allocation to competition gives males greater fitness than offspring care for a range of circumstances that are dependent on life-history parameters and, for the large-game hunting application, frequency dependent. The greater the collective benefit, the more individuals can be selected to supply it. 
\end{abstract}


\pagebreak


\section{Introduction}\label{sec:intro}

Parental care and mating competition are two components of reproductive effort that offer distinct fitness-related payoffs. The first of these reproductive strategies, parental care, contributes to the differential welfare or reproductive success of current offspring, here indexed by differential survivorship. A separate path, mating competition, contributes to the chance of additional conceptions. 

Both sexes are faced with an allocation problem---does one invest reproductive effort into parental care or into mating competition \citep{Houston05, Klug12, McNamara00, Parker02, Trivers72}? Observations of this problem by many early commentators \citep{Bateman48,Darwin59,Darwin74, Fisher30,Trivers72} found that in a wide range of taxa, including many mammalian species, males tend more towards competing for fertilisation opportunities than do females \citep[see][for historical review]{Dewsbury05}, who tend to allocate more effort to parental care \citep{Balshine12,CluttonBrock91,Cockburn06,Kokko12_sex, Queller97}. 

The male-specific payoffs to care and competition have driven males to more commonly display competitive traits. In this study we focus on mammals, whose internal fertilisation leads to increased precopulatory sexual selection, which in turn cascades to the evolution of Darwinian sex roles \citep{Janicke16,Parker14b}. Selection can favour males that reduce their effort into care in response to the tradeoff between traits that improve offspring survival and competitive traits that improve conception chances and mating success \citep{Andersson94,CluttonBrock91, CluttonBrock92, Parker15, Queller97}. This tradeoff is a consequence of the ``Fisher condition'' \citep{Fisher30} and other anisogamy-related differences \citep{CluttonBrock92,Lehtonen14,Lehtonen16,Parker72,Parker14a} that define the payoffs of mate competition and paternal care given that investment in each strategy results in either immediate marginal payoffs to differential survivorship of offspring or an increase in paternities obtained. 

Here, we present a simple mathematical representation of this payoff structure of male reproductive effort, given that males remain in the mating pool regardless of their parental status \citep[time in or out of mating pool as conceptualised by in][]{CluttonBrock92}. Our mathematical model explores the two separate pathways of care and competition and the subsequent equilibrium strategy, given the payoffs each strategy confers. The ordinary differential equation (ODE) model explores how investment in traits that improve differential offspring survivorship or increase the number of obtained paternities shapes the persistent strategy at equilibrium. In previous studies we have shown how male-biased sex ratios \citep[the bias discussed in][]{Coxworth15} favour one kind of mating competition (mate guarding) over multiple mating (another form of mating competition) and paternal care \citep{Loo17a, Loo17b,Rose19}. Others have similarly demonstrated the effect of female availability on the establishment of competitive strategies \citep{Schacht16}. In contrast, \citet{Kokko08b} had previously predicted greater proportions of care in male-biased populations, though this claim was later corrected \citep{Fromhage16}. \possessivecite{Jennions17} modelling found no causal role for sex ratio in the evolution of paternal care. Since this effect has been considered extensively, here we ignore these effects of adult sex ratio and instead consider how \emph{only} changes to immediate marginal payoffs can drive different mammalian male strategies.

We then apply this model to the problem of hunter-gatherer large-game hunting by modifying the payoff structure so that it incorporates both survival benefits directed to offspring, as well as survival benefits provided to all by hunters that net the provider a mating advantage. Selection at the level of the individual could favour providing collective foods if hunters earned individual benefits for all in the group. With this model we aim to contribute to the ongoing discussion regarding whether large-game hunting is a form of paternal care or a means by which males compete for mating opportunities.

The problem of hunter-gatherer large-game hunting stems from observations that hunting men usually procure large-game meat while women usually gather plant food \citep[cross cultural comparisons explored in][]{bird1999cooperation, marlowe2007hunting}. This has been observed in the Hiwi in Venezuela \citep{Hawkes91a,hill1985men,hill1987foraging,  hurtado1985female, hurtado1992trade}, the Ache in Paraguay \citep{Hawkes91a,hill1985men,hill1987foraging,  hurtado1985female, hurtado1992trade}, and the Hadza in Tanzania \citep{berbesque2009sex, Hawkes91b,hawkes1997hadza,Hawkes01,hawkes2001hunting}. Popular accounts attribute this pattern to a sexual division of labour adopted by couples to cooperatively care for their children via the provision of food. This cooperative parental investment is often viewed as a unique human behaviour compared to other primates \citep{Lancaster83}, and male investment into feeding themselves and their families has been proposed as a key driver of human evolution \citep{isaac1978food, Washburn68}. 

However, contrary to suggestions of cooperative care, observations also consistently show that acquired large game is consumed by many. Distributions are not under the hunter's control; they provide a survival benefit to all consumers. Subsequent familial consumption rates do not correlate with individual large-game hunting success \citep{Hawkes93, Hawkes01}---such hunting behaviour does not align with suggestions of paternal care. 

To explain this, the show-off hypothesis proposed that males hunt as a form of costly display facilitating male mate competition, rather than paternal care \citep{Hawkes91b, Hawkes93,Hawkes14}. Agents should contribute to the supply of the community public good only if there is a benefit to the supplier that outweighs the benefit from any foregone opportunities, such as constant, directed provision of food or care to direct offspring. Thus, the question of hunting big game that is inevitably shared (the common good) versus small game that is retained arises. Hadza hunters would obtain more for their households by pursuing frequently attainable small game rather than passing these up to continue searching for shared large game \citep{Hawkes91b}. 

This observation suggests that the pursuit of small game would result in greater benefit, \emph{if} the framework for their decision making is solely household provisioning and consumption. Since small-game hunting is rarely a focus of Hadza men and large-game hunting has high failure rates and rare bonanzas go mostly to others, \citet[pp. 341]{Hawkes93} hypothesised that ``the incentive for providing widely shared goods is favourable attention from other group members", i.e., a competitive benefit over other competing males (either by making hunters desirable collaborators or dangerous opponents) through an honest signal of quality \citep{Grafen90a, Grafen90b,Hawkes02,Hawkes14,Zahavi75,Zahavi77,Zahavi91,Zahavi95}. Large-game hunting investment increases hunter's paternity chances---hunters have higher reproductive success not because their offspring have higher survival but because they have more of them \citep[quantified in][]{Jones16}. This arena of signalling is of special interest to the audience because they receive consumption benefits in addition to information regarding male qualities. Evidence from the Ache people in Paraguay further supports this argument \citep{Hawkes91a,Hill96}, with better Ache hunters having higher reproductive success. 

Our modified ODE model considers whether forms of mating competition that supply benefits to all (which may take a number of alternative forms, e.g., predator defence) affect tradeoffs with paternal care. As in the simple model, caring males gain benefits through differential survivorship of their offspring by acquiring resources consumed entirely by these offspring. Competers (or hunters) forgo these survival benefits for their own offspring to invest in large-game hunting that pays off in increased relative number of paternities and additionally provides a survival benefit to all. We show that mating competition through large-game hunting can be the dominant stable reproductive strategy over a wide range of parameter values, including cases where care ensures the survival of offspring, and further demonstrate a region of frequency dependence within which either large-game hunting or caring can persist dependent on the composition of the starting population.

By establishing a model within which payoffs can be easily defined we show a range of situations within which the benefits of competition outweigh those of care, \emph{even} when care ensures juvenile survival. However, our model also shows regions where care outperforms competition, given certain life-history, offspring benefit parameters, and initial population proportions. This simple dynamical system provides a guide to investigating problems of allocating reproductive effort and demonstrates the strength of payoffs to competition over care in males even when care ensures offspring survival (i.e., all offspring of carers necessarily survive to reproductive age).

\section{Model of paternal care and mating competition} \label{sec:model}

 \subsection{Populations}

We first formulate a simple and easily generalisable ODE model, starting with a population comprised of males and females. To investigate the persistence of paternal care and mating competition given a simple payoff structure, we compare two pure strategies: males who invest all of their time and effort in paternal care, and those who invest in mating competition. Mating competition can take a variety of forms, such as contests, attractiveness to females, or scrambles. For the sake of simplicity and generalisation this compartment is defined as males with any competitive trait leading to increased likelihood of paternity. These strategies are defined by the variables $C$ for paternal carers and $M$ for competing males. The female population is denoted by the variable $F$. All populations are considered to be adult, fertile populations.   

\subsection{Ordinary differential equation system}

To determine the equilibrium behaviours of each strategy and the winning strategy given different parameters, we formulate a system of ODEs for the three populations, $C$, $M$ and $F$. This system is given by
\begin{eqnarray}
\text{Carers  } & \mathlarger{\frac{dC}{dt}} &=\,\,\, \underbrace{\frac{b}{2} \overbrace{\exp\left(-(1-c)\tau\mu\right)}^{\substack{\text{Offspring survival} \\ \text{fraction for carers}}} F \overbrace{\frac{C}{C+\alpha M}}^{\substack{\text{Fraction of} \\ \text{paternities won} \\ \text{by carers}}} } _{\text{Birth term}} - \overbrace{(\mu_m + \epsilon T) C}^{\substack{\text{Density-dependent}\\ \text{death rate}}}  ,\label{eqn:c1.1} \\ 
\text{Competers  } & \mathlarger{\frac{dM}{dt}} &=\,\,\, \underbrace{\frac{b}{2} \overbrace{\exp \left( -\tau\mu \right)}^{\substack{\text{Offspring survival}\\\text{fraction for competers}}} F \overbrace{\frac{\alpha M}{C+\alpha M}}^{\substack{\text{Fraction of} \\ \text{paternities won} \\ \text{by competers}}}}_{\text{Birth term}} - \overbrace{(\mu_m + \epsilon T) M}^{\substack{\text{Density-dependent}\\ \text{death rate}}} , \label{eqn:m1.1}\\
\text{Females  } & \mathlarger{\frac{dF}{dt}} &=\,\,\, \underbrace{\frac{b}{2} \frac{C \exp\left( -(1-c)\tau\mu \right) + \alpha M \exp\left( -\tau\mu \right)}{C+\alpha M} F}_{\text{Birth term}} - \overbrace{(\mu_m + \epsilon T) F.}^{\substack{\text{Density-dependent}\\ \text{death rate}}} \label{eqn:fem1.1}
\end{eqnarray}

These equations represent the birth and death of each population and implicitly model the transition from childhood to maturity. The differential equation given by Equation~\eqref{eqn:c1.1} gives the change to the population of carers over time. The first term in this equation expresses birth, which is itself made up of two components. The first component gives the offspring survival fraction for the offspring of carers. Assuming that females have an intrinsic birth rate of $b$, we assume that offspring survival follows an exponential decay (a Poisson process). Thus, the surviving fraction of the offspring of carers that reach the age of maturity to fertile adulthood $\tau$ takes an exponential form given by $\exp(-(1-c)\tau\mu)$. Here, the mortality rate for these offspring is $(1-c)\mu$, where $\mu$ is the baseline juvenile mortality rate and $c$ is the survival benefit provided to the offspring of carers. When $c=1$, mortality of the offspring of carers is $0$, meaning that all offspring survive to maturity, and when $c=0$, mortality is equal to that of the offspring of competing males, $\mu$.

The second component of the birth term is the proportion of paternities obtained (or won) by the population of caring males over competers $\frac{C}{C+\alpha M}$. Here the denominator is the sum of males weighted by their competitive standing; $\alpha$ represents competers' relative success in mating competition. 

The second term in the carers' differential equation (and similarly for all differential equations in the system) gives the death rate of carers. We presume the adult mortality rate is given by $ \mu_m + \epsilon T$, where $\mu_m$ the baseline adult mortality rate, $T = F+C+M$ is the total population size and $\epsilon$ is small. This density-dependent term causes the population to settle at a finite equilibrium. The adult death rate is the same for both other compartments---competers and females.

The differential equation for competers (Equation~\eqref{eqn:m1.1}) follows the same birth-death structure. For these competing males, offspring mortality is given by the baseline juvenile mortality $\mu$ with no survival benefit from care. This corresponds to an offspring survival fraction of $\exp(-\tau \mu)$. Here, the proportion of paternities obtained by competing males is given by $\frac{\alpha M}{C+\alpha M}$; $\alpha$ is the paternity advantage apportioned to competers relative to carers. We restrict $\alpha>1$ to model the payoff to competition through increased paternities. For example, if $\alpha=2$, each competing male is twice as likely to obtain a paternity as one of his caring counterparts. The weighting parameter $\alpha$ can theoretically be any value greater than 1; however, to consider realistic values, we investigate $1\leq \alpha \leq 2$, so that the weighting advantage does not exceed 2-to-1. 
Furthermore, it is assumed that half of the surviving offspring are male and half are female. Thus, all birth terms are multiplied by a factor of $1/2$. 

Payoffs to each strategy are modelled as reproductive advantages, where care pays off through a greater proportion of offspring surviving to adulthood (as captured by the difference between the offspring survival fractions of each strategy), and competition pays off through winning greater relative numbers of paternities. Defining benefits in this way remains consistent with Darwin's definition of sexual selection by capturing the notion that advantages of mating accrue through \emph{immediate marginal gains} in conceptions. It is the balance between competition (investment in $\alpha$) and care (investment in survival benefit $c$), and their subsequent effects on strategies that we are interested in here. 

Given initial conditions such that the population has equal sex ratio, i.e., $F_0 =C_0+M_0$, the sex ratio remains equal for any given time, i.e., $F=C+M$, for all $t$. Under this assumption, the differential equation for the rate of change of the female population, Equation (\ref{eqn:fem1.1}), can be disregarded and the system reduced to a two-dimensional system in $C$ and $M$. The simplified model is given by
\begin{eqnarray}
\begin{aligned}
\frac{dC}{dt} &= \frac{b}{2} \exp\left( -(1-c)\tau\mu \right) (C+M) \frac{C}{C+ \alpha M} - (\mu_m + \epsilon T) C ,   \\
\frac{dM}{dt} &= \frac{b}{2} \exp \left( - \tau \mu \right) (C+M) \frac{\alpha M}{C+ \alpha M} - (\mu_m + \epsilon T) M , 
\end{aligned}\label{eqn:simpleODE}
\end{eqnarray}
with parameters defined as above, and where $T = F+C+M = 2(C+M)$.

\subsection{Results}

With this two-dimensional model, analytic steady-state analysis can be performed. This highlights the parameter regions and conditions within which each strategy outperforms the other, giving regions where natural selection will favour traits that increase payoffs to either mating competition or parental care. Two steady states that correspond to equilibria where either the strategy of care wins (SS Care) or where mating competition wins (SS Competition) are given by
\begin{eqnarray*}\begin{aligned}
\text{SS Care} &&  (C_1,M_1) &= \left( \frac{b \exp (-(1-c)\tau \mu) - 2 \mu_m }{4 \epsilon},0 \right) , \\
\text{SS Competition} && (C_2,M_2) &= \left( 0,\frac{b \exp (-\tau\mu) - 2 \mu_m }{4 \epsilon} \right) .
\end{aligned}\end{eqnarray*}
Stability analysis then gives that care (SS Care) persists when the following two conditions hold:
\begin{eqnarray}
\mu_m < \frac{b}{2}\exp(-(1-c)\tau\mu), \label{eqn:Cstab1}\\
\alpha < \exp(c\tau\mu) \label{eqn:Cstab2}.
\end{eqnarray}
Conversely, the steady state where competitive behaviours persist (SS Competition) is stable when the following conditions hold:
\begin{eqnarray}
\mu_m < \frac{b}{2}\exp(-\tau\mu), \label{eqn:Mstab1}\\
\alpha > \exp(c\tau \mu)\label{eqn:Mstab2}.
\end{eqnarray}
Inequalities~\eqref{eqn:Cstab1} and~\eqref{eqn:Mstab1} are satisfied when birth rates exceed the death rates of populations $C$ and $M$ respectively at those steady states. The right hand sides of these inequalities $\exp(-(1-c)\tau\mu)$ and $\exp(-\tau\mu)$ are the fractions of $C$ and $M$ respectively who survive to reproductive age. 

Inequality~\eqref{eqn:Cstab2} is the condition for the birth rates of $C$ to exceed $M$ and, similarly, inequality~\eqref{eqn:Mstab2} for the birth rate of $M$ to exceed $C$. These stability conditions are the converse of each other and therefore, as life-history parameters and payoffs change, the equilibrium solution of the ODE system will jump from one steady state to the other, given that the other conditions for stability are satisfied. It is this inequality that drives stability. Thus, it is evident that stability is dependent on life-history parameters, $\mu$, $\mu_m$, $\tau$, and $b$, and not upon the (initial) relative frequency of caring and competing strategies; steady states are frequency independent. If the per capita birth rate of one group exceeds that of the other, that group will ultimately dominate the population, and we do not expect both strategies to occur simultaneously. 

As a representative life-history, we focus on humans and the evolution of competition and care in our human lineage. Thus, we investigate regions of stability for baseline hunter-gatherer life-history parameters, $b=0.3$ year$^{-1}$, $\tau=18$ years, $\mu = 1/30$ and $\mu_m=1/40$ year$^{-1}$. This female birth rate, $b = 0.3$, corresponds to birth intervals of approximately $1/0.3 \sim 3$ years, allowing an approximate $1$ year for female conception and delivery, with an additional $2$ years for weaning (until toddlers can be nutritionally independent of mothers) \citep{Kim12, Kim14,Sear08}. Further, our approximations for the age of maturity, $\tau$, and adult mortality rate, $\mu_m$, are informed by \cite{Kim14}, who scale their life-history parameters by life expectancy and who are themselves informed by demographic data from great apes and humans \citep{Gurven07,Knott01,Robbins06,Robson06,Sellen07}. This scaling originally established in \cite{Kim12} corresponds to $\mu_m = 1/L$ and $\tau = L/2.5 + \tau_0$, where $L$ is life expectancy and $\tau_0$ is a constant age of weaning. We note that human age-specific mortality, like that of other mammals, is high at birth, dropping until near age-at-maturity and then climbing again in adulthood \citep{Gurven07,volk2013infant}. We assume a higher juvenile mortality rate $\mu$ and then a constant adult mortality $\mu_m$, for simplicity.  

\begin{figure}[ht!]
\begin{center}
\vspace{10pt}
\includegraphics[scale=0.6]{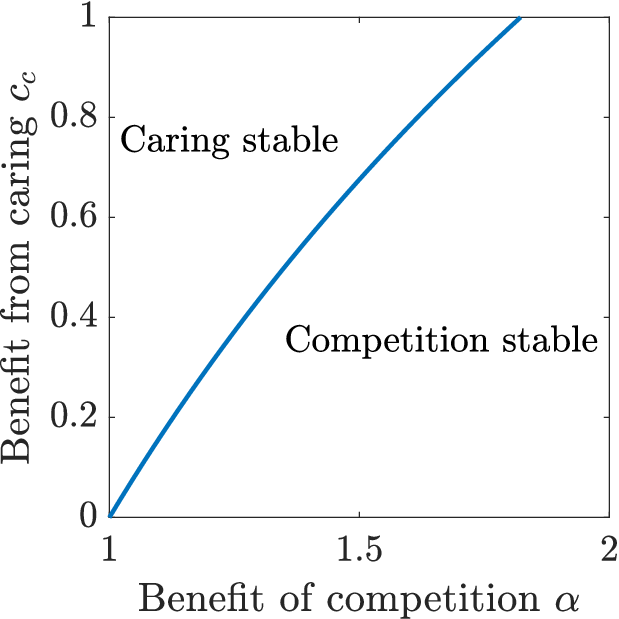}
\end{center}
\internallinenumbers\caption{Bifurcation obtained from the analytic solution and steady states given in Equations~\eqref{eqn:Cstab1}-\eqref{eqn:Mstab2}, in parameters of competition, $\alpha$, and care, $c$. This determines which equilibrium is stable and thus, which strategy overcomes the other. To the right of the line, mating competition outperforms the caring strategy, and to the left of the line, care persists over mating competition. The equation describing the bifurcation curve is $c = \ln(\alpha)/(\tau\mu)$, derived from Equations~\eqref{eqn:Cstab2} and~\eqref{eqn:Mstab2}. Baseline parameter choices are given in Table~\ref{table:Ch4_ODEparams}.}
  \label{fig:bif1}
\end{figure}

For the human-like values, stability conditions given by Equations \eqref{eqn:Cstab1} and \eqref{eqn:Mstab1} hold true. Thus, stability is driven by the values of $\alpha$ and $c$ as given by Equations \eqref{eqn:Cstab2} and \eqref{eqn:Mstab2}. For this hunter-gatherer example, the inequality determining which equilibrium is stable (i.e., whether inequality \eqref{eqn:Cstab2} or \eqref{eqn:Mstab2} is satisfied) in the $\alpha$-$c$ plane is shown in Figure~\ref{fig:bif1}. We describe this inequality as a bifurcation.

To the left of the bifurcation in Figure~\ref{fig:bif1} the caring strategy wins, while to the right side of the bifurcation mating competition wins, overcoming the caring population. When males have increased relative paternity benefits from competition, i.e., $\alpha$ is slightly greater than $1$, mating competition can outperform paternal care. For example when $c = 0.25$ (such that the offspring of carers have a 25\% higher survival rate per year they are juveniles) mating competition outperforms care when $\alpha >1.12$. In the extreme and unrealistic case where the benefit of paternal care is maximised, $c=1$, and all the offspring of carers reach maturity, the relative reproductive benefit need only be $\alpha > 1.6 $ in order for mating competition to outperform care. That is, if mating competition causes competitors to be $1.6$ times as likely to obtain paternities as paternal carers, despite care resulting in the survival of all offspring to adulthood, competition still outperforms care. 

\begin{table}[htp!] 
\begin{center}
	\caption[Summary of baseline parameters used in the ODE model comparing payoffs to competition and care.]{Summary of baseline parameter estimates used in analysing and simulating the ODE system. Simulations and analysis use these baseline parameters, based on hunter-gatherers \citep{Gurven07,Kim12,Kim14,Robson06,Sear08,Sellen07}, unless otherwise stated. }\label{table:Ch4_ODEparams}
\renewcommand*{\arraystretch}{1.5}
\begin{tabular}{  ccc} 
\textbf{Parameter} & \textbf{Interpretation}  & \textbf{Value} \\
\hline
$b$	& Birth rate	&	$0.3$ year$^{-1}$\\ 
$\tau$	& Age of independence and transition to fertility 	& $18$	\\
$\mu_m$	& Adult mortality rate				&			$1/40$ \\
$\mu$	& Baseline juvenile mortality rate				&			$1/30$ \\
$c$	& Survival benefit to offspring of caring males 		&	$0<c<1$	\\	
$\alpha$	& Marginal reproductive benefit of mating competition	&	$\alpha \geq 1$	\\ 
$M_0$ & Initial competing male population & \\
$C_0$ & Initial caring male population & \\
\end{tabular}
\end{center}
\end{table}

The payoff structure of the two-strategy ODE model given by System~\eqref{eqn:simpleODE} can be adapted to include additional benefits to each strategy and similar steady-state analysis performed to investigate the regions within which either strategy outperforms the other. In the following section we show an example of this, where the payoff structure of the model is modified to include survival benefits to all via the provision of a consumable community good by competers.


\section{Large-game hunting as a public good}\label{sec:publicgood}

The simple ODE model given by Equation~\eqref{eqn:simpleODE} in Section~\ref{sec:model} assumed that benefits to offspring survival came directly and solely from a caring male and offspring mortality was adjusted by a constant proportion $c$. However, as was noted in the introduction, the question of why men hunt large game instead of providing a more consistent resource via small game or gathering plant foods is emblematic of the larger question of competition versus care.

Competition manifests in large-game hunting behaviours that usually fail, but occasional bonanzas are widely distributed -- game is \emph{not} differentially supplied to the hunter's own offspring. We model this with following system:
\begin{equation} 
\begin{aligned}\label{eqn:publicgoodssystem}
\mathlarger{\frac{dC}{dt}} &= \underbrace{\frac{b}{2} \overbrace{\exp\left(-\left(1-c_h \frac{M}{M+C} \right)(1-c_c)\tau\mu\right)}^{\substack{\text{Offspring survival} \\ \text{fraction for carers}}} (C+M) \overbrace{\frac{C}{C+\alpha M}}^{\substack{\text{Fraction of} \\ \text{paternities won} \\ \text{by carers}}} } _{\text{Birth rate}} - \overbrace{(\mu_m + \epsilon T) C}^{\substack{\text{Density-dependent}\\ \text{death rate}}}  , \\ 
\mathlarger{\frac{dM}{dt}} &= \underbrace{\frac{b}{2} \overbrace{\exp \left( -\left(1-c_h \frac{M}{M+C}\right)\tau\mu  \right)}^{\substack{\text{Offspring survival}\\\text{fraction for competers}}} (C+M) \overbrace{\frac{\alpha M}{C+\alpha M}}^{\substack{\text{Fraction of} \\ \text{paternities won}\\ \text{by competers}}}}_{\text{Birth rate}} - \overbrace{(\mu_m + \epsilon T) M}^{\substack{\text{Density-dependent}\\ \text{death rate}}} . 
\end{aligned} 
\end{equation}
Here, the first term in each differential equation again refers to the offspring survival fraction for that strategy. For carers this is given by  
\begin{eqnarray*}
    \exp \left( -\left( 1-c_h \frac{M}{M+C} \right) (1-c_c)\tau\mu \right),
\end{eqnarray*}
assuming that death of juveniles follows an exponential decay, as in the previous model. We define the mortality of the offspring of carers as $\left( 1 - c_h \frac{M}{M+C} \right) (1-c_c) \mu$, where $c_h$ is the benefit of large-game, $c_c$ is the survival benefit provided by carers, and $\mu$ is baseline juvenile mortality, with male populations of large-game hunters $M$ and carers $C$. We assume the survival benefit provided by hunters and shared amongst all is inversely proportional to the total number of males in the population, representing the share of acquired game provided to each male and his family. Thus, the benefit $c_h$ is multiplied by the fraction $\frac{M}{M+C}$. The survival benefit directed to the offspring of carers $c_c$ is equivalent to the parameter $c$ in the model from Section~\ref{sec:model} and defines the direct offspring survival benefit from care.

In the same way, the surviving fraction of the offspring of hunters is 
\begin{eqnarray*} 
\exp\left( -\left( 1 - c_h \frac{M}{M+C} \right)\tau\mu \right),
\end{eqnarray*}
with parameters as above. This is based on a mortality rate $\left( 1 - c_h \frac{M}{M+C} \right) \mu$ for the offspring of hunters. 

The fractions of paternities obtained by carers and competers take the same form as in System~\eqref{eqn:simpleODE}, and adult mortality takes a density-dependent form given by the second terms of both equations in System~\eqref{eqn:publicgoodssystem}.

To simplify analysis, we reduce System~\eqref{eqn:publicgoodssystem} to a one-dimensional system. We set $\epsilon=0$ and normalise the populations by defining $u = M/(M+C)$, the frequency of hunters in the males-only population. Thus, the frequency of carers is $C/(M+C) = 1-u$. We obtain
\begin{linenomath*}
\begin{align}
    \dfrac{du}{dt} &= \dfrac{\dfrac{dM}{dt}}{M+C} - \dfrac{M\dfrac{d}{dt}(M+C)}{(M+C)^2} \notag\\
     &= \frac{b}{2}\dfrac{\exp \left(-\left(1-c_h u\right)\tau\mu\right)}{1 + (\alpha-1) u}f(u)g(u),\label{eqn:udot}
\end{align}\end{linenomath*}
with \begin{eqnarray*}f(u) &=& \alpha - \exp\left(c_c\left(1-c_h u\right)\tau\mu\right) \\ &=& \alpha - \exp\left(c_c\tau\mu\right)\exp\left(-c_hc_c\tau\mu u\right)\end{eqnarray*} 
and \begin{eqnarray*}g(u) &=& u(1-u).\end{eqnarray*}
Note that $f(u)$ is an exponential function that approaches $\alpha$ asymptotically from below as $u\to\infty$. Thus it always has one real root, or solution. Further, $g(u)$ is a concave down parabola whose roots are $u=0$ and $u=1$, corresponding to populations of carers only or of competers only, respectively. Observe that the adult mortality rate $\mu_m$ drops out of this simplified model altogether, since both populations are subject to the same density-dependent death rate.

This one-dimensional representation assumes that the birth rate is sufficient for the population to persist, regardless of the strategic composition of the population. Reducing to this one-dimensional model reduces the number of conditions to consider in stability analysis. We note that we can obtain steady states for our original two-dimensional System~\eqref{eqn:publicgoodssystem}. When converted to population densities they correspond exactly to the easily obtained steady states of Equation~\eqref{eqn:udot}. Thus, we analyse our reduced model as follows.

\subsection{Results}

Equilibrium solutions to the differential equation~\eqref{eqn:udot} correspond to $g(u) = 0$ or $f(u) = 0$. The solutions to $g(u) = 0$ are $u=0$ and $u=1$, corresponding to a population of only carers and a population of only competers respectively. Observe that $f(u)$ is an exponential function that approaches $\alpha$ from below. Thus it must have one real root, given by 
\begin{eqnarray*}
u^* &=& \frac{1}{c_h}\left(1 - \frac{\ln\alpha}{c_c\tau\mu}\right).
\end{eqnarray*}
When $u^*\in(0,1)$ this represents a physical coexistence equilibrium; either strategy may be stable.

Observe that $du/dt>0$ for $-1/(\alpha-1)<u<\operatorname{min}\{u^*,0\}$, and $du/dt<0$ for $u>\operatorname{max}\{u^*,1\}$. Since $du/dt$ is continuous for $u>-1/(\alpha-1)$ and has three roots, stability analysis indicates that the left-most and right-most roots are stable steady states, and the middle root, whichever one it is, is unstable. Thus, depending on the value of $u^*$, we have three possible cases (i.e., if $u^*<0$, $u^*>1$ or $0<u^*<1$). Stability is frequency dependent. We illustrate this in Figure~\ref{fig:uequilibria}, and outline each case below.

When $u^*<0$, illustrated in Figure~\ref{fig:u1stab}, we have $u=0$ is unstable and $u=1$ is stable. Any physically realistic initial condition $u(0)\in(0,1)$ will ultimately approach $u=1$; hunting will succeed. The only way that caring can persist is if the initial frequency of hunters is $u(0) = 0$. The condition for $u=1$ to be stable is given by $\exp(c_c\tau\mu) < \alpha$, such that the mating benefit due to competition is always greater than the survival benefit due to provisioning, even in the absence of shared public goods.

\begin{figure}
    \centering
    \begin{subfigure}{0.45\textwidth}
        \centering\begin{tikzpicture}[>=stealth',auto,thick,scale=2.5,every node/.style={scale=1.3}]

\tikzmath{real \xaxmin; \xaxmin = -0.45; real \xaxmax; \xaxmax = 1.15; real \yaxmin; \yaxmin=-0.5; real \yaxmax; \yaxmax=0.5; real \b; \b = 0.3; real \al; \al = 1.1; real \t; \t = 18; real \m; \m = {1/30}; real \cc; \cc = 0.15; real \ch; \ch = 0.2; real \ustar;}

\draw[ultra thick,domain=\xaxmin:\xaxmax,samples=32] plot ({\x},{1000*\b/2*exp(-(1-\ch*\x)*\t*\m)/(1+(\al-1)*\x)*\x*(1-\x)*(\al-exp((1-\ch*\x)*\cc*\t*\m))});
\draw[->] (-0.5,0) -- (1.5,0) node[below] {\scalebox{1}{$u$}};
\draw[->] (0,-0.5) -- (0,0.5) node[right] {$\frac{du}{dt}$};
\draw[->,thick] (0,0) to (1-0.05,0);
\draw[fill=white,thick] (0,0) circle (0.75pt) node[below right] {$0$};
\draw[fill=black] (1,0) circle (0.75pt) node[below left] {$1$};
\draw[fill=black] ({(1 - ln(\al)/(\cc*\t*\m))/\ch},0) circle (0.75pt) node[below] {$u^*$};

\end{tikzpicture}
    \caption{$c_c = 0.15$, $c_h = 0.2$, such that the competition-only equilibrium is stable, with $u^*<0$. All initial conditions $u(0)\in(0,1)$ will approach $1$, as $du/dt>0$ for $0<u<1$.}
    \label{fig:u1stab}
        
    \end{subfigure}
    \hspace{30pt}
    \begin{subfigure}{0.45\textwidth}
        \centering
    \begin{tikzpicture}[>=stealth',auto,thick,scale=2.5,every node/.style={scale=1.3}]

\tikzmath{real \xaxmin; \xaxmin = -0.13; real \xaxmax; \xaxmax = 1.45; real \yaxmin; \yaxmin=-0.5; real \yaxmax; \yaxmax=1; real \b; \b = 0.3; real \al; \al = 1.1; real \t; \t = 18; real \m; \m = {1/30}; real \cc; \cc = 0.215; real \ch; \ch = 0.2; real \ustar;}

\draw[ultra thick,domain=\xaxmin:\xaxmax,samples=32] plot ({\x},{1000*\b/2*exp(-(1-\ch*\x)*\t*\m)/(1+(\al-1)*\x)*\x*(1-\x)*(\al-exp((1-\ch*\x)*\cc*\t*\m))});
\draw[->] (-0.5,0) -- (1.5,0) node[below] {$u$};
\draw[->] (0,-0.5) -- (0,0.5) node[right] {$\frac{du}{dt}$};
\draw[->,thick] (1,0) to (0.05,0);
\draw[fill=black] (0,0) circle (0.75pt) node[below left] {$0$};
\draw[fill=white,thick] (1,0) circle (0.75pt) node[below] {$1$};
\draw[fill=black] ({(1 - ln(\al)/(\cc*\t*\m))/\ch},0) circle (0.75pt) node[below] {$u^*$};

\end{tikzpicture}
    \caption{$c_c = 0.215$, $c_h = 0.2$, such that the caring-only equilibrium is stable, with $u^*>1$. All initial conditions $u(0)\in(0,1)$ will approach $0$, as $du/dt<0$ for $0<u<1$.}
    \label{fig:u0stab}
        
    \end{subfigure}\\
    \begin{subfigure}{0.45\textwidth}
        \centering
    \begin{tikzpicture}[>=stealth',auto,thick,scale=2.5,every node/.style={scale=1.3}]

\tikzmath{real \xaxmin; \xaxmin = -0.235; real \xaxmax; \xaxmax = 1.275; real \yaxmin; \yaxmin=-0.5; real \yaxmax; \yaxmax=1; real \b; \b = 0.3; real \al; \al = 1.1; real \t; \t = 18; real \m; \m = {1/30}; real \cc; \cc = 0.18; real \ch; \ch = 0.2; real \ustar;}

\draw[ultra thick,domain=\xaxmin:\xaxmax,samples=32] plot ({\x},{1000*\b/2*exp(-(1-\ch*\x)*\t*\m)/(1+(\al-1)*\x)*\x*(1-\x)*(\al-exp((1-\ch*\x)*\cc*\t*\m))});
\draw[->] (-0.5,0) -- (1.5,0) node[below] {$u$};
\draw[->] (0,-0.5) -- (0,0.5) node[right] {$\frac{du}{dt}$};
\draw[->,thick] ({(1 - ln(\al)/(\cc*\t*\m))/\ch},0) to (1-0.05,0);
\draw[->,thick] ({(1 - ln(\al)/(\cc*\t*\m))/\ch},0) to (0.05,0);
\draw[fill=black] (0,0) circle (0.75pt) node[below left] {$0$};
\draw[fill=black] (1,0) circle (0.75pt) node[below] {$1$};
\draw[fill=white,thick] ({(1 - ln(\al)/(\cc*\t*\m))/\ch},0) circle (0.75pt) node[below] {$u^*$};

\end{tikzpicture}
    \caption{$c_c = 0.18$, $c_h = 0.2$, such that both $u=0$ and $u=1$ are stable, with $0<u^*<1$. An initial condition $u(0)\in(0,u^*)$ will approach $0$, and an initial condition $u(0)\in(u^*,1)$ will approach $1$, as $du/dt<0$ for $0<u<u^*$, and $du/dt>0$ for $u^*<u<1$.}
    \label{fig:bothstab}
        
    \end{subfigure}
    \caption{Graphs of Equation~\eqref{eqn:udot} for baseline parameters and $\alpha = 1.1$,  with~(\subref{fig:u1stab}) $c_c = 0.2$, $c_h = 0.2$,~(\subref{fig:u0stab}) $c_c = 0.25$, $c_h = 0.12$, and~(\subref{fig:bothstab}) $c_c = 0.25$, $c_h = 0.25$. The middle steady state is always unstable, indicated by a hollow dot; stable steady states are indicated by solid dots. Flow directions for $u\in[0,1]$ are indicated by arrows on the $u$-axis.}
    \label{fig:uequilibria}
\end{figure}

When $u^*>1$, illustrated in Figure~\ref{fig:u0stab}, we have $u=0$ is stable and $u=1$ is unstable. Any initial condition $u(0)\in(0,1)$ will approach $u=0$. Competition can only persist if the initial frequency of hunters is $u(0)=1$. The condition for $u=0$ to be stable is given by $\exp((1 - c_h)c_c\tau\mu) > \alpha$, such that the survival benefit of provisioning (plus any extra due to shared public goods) always exceeds the mating benefit due to competition.

Figure~\ref{fig:bothstab} gives the case where $0<u^*<1$; the coexistence steady state is physically meaningful (i.e., frequency of competing males is between 0 and 1) and unstable. 
Here, when the frequency of competers in the population is below the critical threshold $u^*$, their birth rate is too low for them to out-compete caring; initial conditions $u(0)\in(0,u^*)$ will result in $u=0$. When initial conditions $u(0)\in(u^*,1)$, the population will approach $u=1$. These results correspond to 
\begin{eqnarray*}\exp((1 - c_h)c_c\tau\mu) < \alpha < \exp(c_c\tau\mu).\end{eqnarray*} 
The mating benefit due to competition $\alpha$ lies between the survival benefits due to provisioning and public goods combined, and provisioning alone.

We do not illustrate the two edge cases, $u^*=0$ and $u^*=1$. These cases give repeated roots of Equation~\eqref{eqn:udot} that do not change the sign of $du/dt$ on the interval $(0,1)$ from the respective cases $u^*<0$ and $u^*>1$. Consequently, the stability conditions can be expanded: $u=1$ alone is stable when $u^*\le0$, and $u=0$ alone is stable when $u^*\ge1$, and both $u=0$, $1$ are stable when $0<u^*<1$. When both $u=0$ and $1$ are stable, the winning strategy is dependent the initial frequency of each strategy.

\begin{figure}[ht!]
\begin{center}
\begin{subfigure}[b]{0.6\textwidth}
    \begin{center}
    \includegraphics[width=\textwidth]{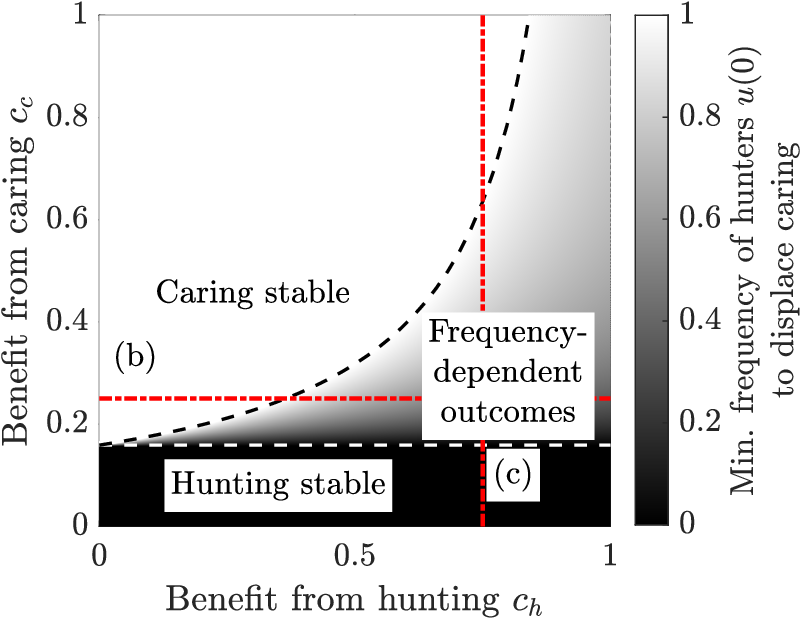}
    \caption{Entire $c_h c_c$-plane for $\alpha = 1.1$.}
    \label{subfig:chccplane}
    \end{center}
\end{subfigure}
\begin{subfigure}[b]{0.4\textwidth}
    \begin{center}
    \includegraphics[width=\textwidth]{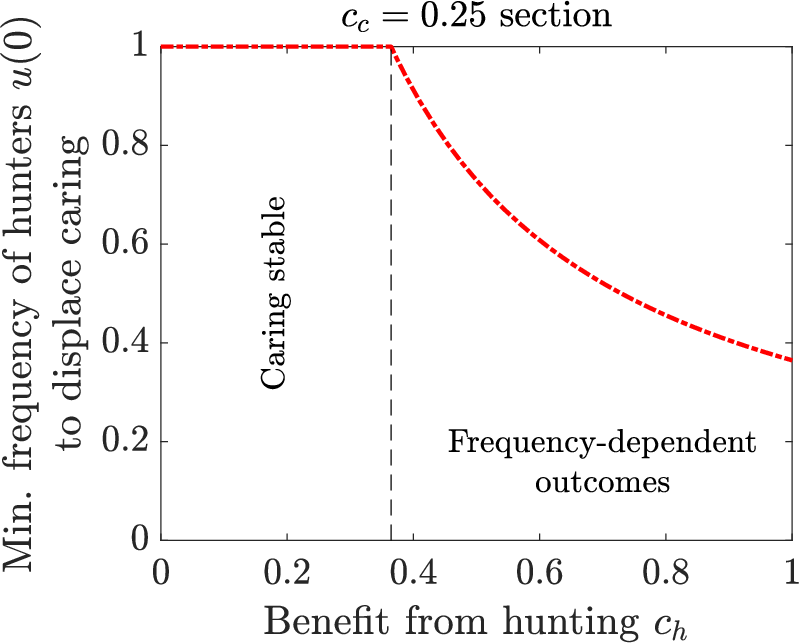}
    \caption{$c_c = 0.25$, $\alpha = 1.1$.}
    \label{subfig:ccdensity}
    \end{center}
\end{subfigure}
\hspace{20pt}
\begin{subfigure}[b]{0.4\textwidth}
    \begin{center}
    \includegraphics[width=\textwidth]{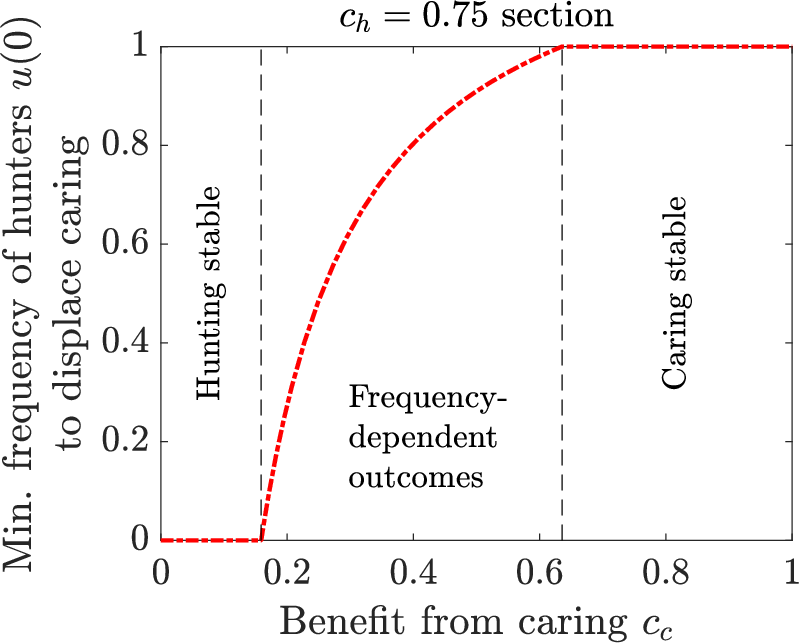}
    \caption{$c_h = 0.75$, $\alpha = 1.1$.}
    \label{subfig:chdensity}
    \end{center}
\end{subfigure}\vspace{-10pt}
\end{center}
	\internallinenumbers\caption{(\subref{subfig:chccplane}) The minimum initial density of hunters $u(0)$ required to displace caring, for $(c_h,c_c)\in[0,1]\times[0,1]$ with $\alpha = 1.1$; white requires an the total initial population to be made up of hunters for hunting to displace care $u(0)=1$, and black corresponds hunting displacing care at any initial frequency $u(0)>0$. 
	The black and white dashed lines show the boundaries between regions where only one strategy is stable, and the region where both strategies may be stable, and are frequency dependent, as given by the scale. The black dashed line is given by $c_c = \frac{\ln\alpha}{(1-c_h)\tau\mu}$, and the white dashed line is $c_c = \frac{\ln\alpha}{\tau\mu}$. The red dash-dot lines in (\subref{subfig:chccplane}) are the corresponding sections in 
	(\subref{subfig:chdensity}) and (\subref{subfig:ccdensity}) of constant $c_c$ and $c_h$ (the horizontal line refers to (\subref{subfig:chdensity}) and the vertical line to (\subref{subfig:ccdensity})).	(\subref{subfig:chdensity}) and (\subref{subfig:ccdensity}) show the minimum initial frequency of hunters to displace caring as red curves (i.e., the equivalent scale in (\subref{subfig:chccplane})). The vertical dashed lines correspond to the dashed lines in (\subref{subfig:chccplane}) that define the frequency-dependent region. Above the red curve hunters will dominate, and below caring will dominate.}
	\label{fig:popsplit}
\end{figure}

To further explore these stability regions, in Figure~\ref{fig:popsplit} we fix the mating benefit of competition $\alpha = 1.1$ and $\tau$, $\mu$ as in Table~\ref{table:Ch4_ODEparams} and plot the region of frequency-dependent outcomes in the $c_h$-$c_c$-plane. Figure~\ref{subfig:chccplane} indicates the minimum initial density of competers $u_0 = M_0/(M_0+C_0)$ required for public goods competition to overcome care by provisioning. The greyscale gradient corresponds to the minimum initial proportion of hunters that will overcome care. The solid black region indicates where $\exp(c_c\tau\mu)<\alpha$, and $u=1$ is stable. Hunting will overcome care. The solid white region indicates where $\exp((1-c_h)c_c\tau\mu)>\alpha$; no invasion of public goods competition can displace provisioning care. The greyscale region between the dashed lines represents the region for which $\exp((1-c_h)c_c\tau\mu)<\alpha<\exp(c_c\tau\mu)$, where the initial frequency of either strategy determines the outcome.

Figures~\ref{subfig:ccdensity} and~\ref{subfig:chdensity} show sections of constant $c_c=0.25$ and $c_h=0.75$ where, if the frequency of hunters exceeds that plotted along the red curve, hunting will dominate. Above the red curve, public goods competers are able to invade a population of carers. 

The existence of a region where the stable equilibrium is frequency dependent suggests a positive feedback loop in the population dynamics: if the density of hunters exceeds the critical threshold given in red in Figure~\ref{subfig:ccdensity} and~\ref{subfig:chdensity} (for given parameters), large-game hunting becomes fitter simply because relatively more of their offspring (both male and female) survive to fertility; conversely, if carers exceed a certain critical density, caring becomes fitter for the same reason. In essence, the relative rate of successful offspring production swings further in favour of the dominant population, whichever side of the critical density the population lies. If the population lies \textit{precisely} on the critical density, the relative rates of offspring production are balanced and the proportion remains the same.

This region where either caring by provisioning or competition by hunting can succeed can be framed in terms of the ``tragedy of the commons" \citep{Hardin68}. Since hunters bring in a shared public good, the presence of carers dilutes the benefit that goes to the offspring of hunters. If the survival benefit to the offspring of carers is high enough to offset the immediate reproductive benefit of large-game hunting, and provided there are enough carers in the population, pursuing large game that is widely shared can become not worthwhile. On the other hand, the corollary existence of a sufficient density of hunters means that their relatively greater success at obtaining paternities can drive caring to extinction.

\begin{figure}[ht!]
\begin{center}
\begin{subfigure}[b]{0.4\textwidth}
    \begin{center}
    \includegraphics[width=\textwidth]{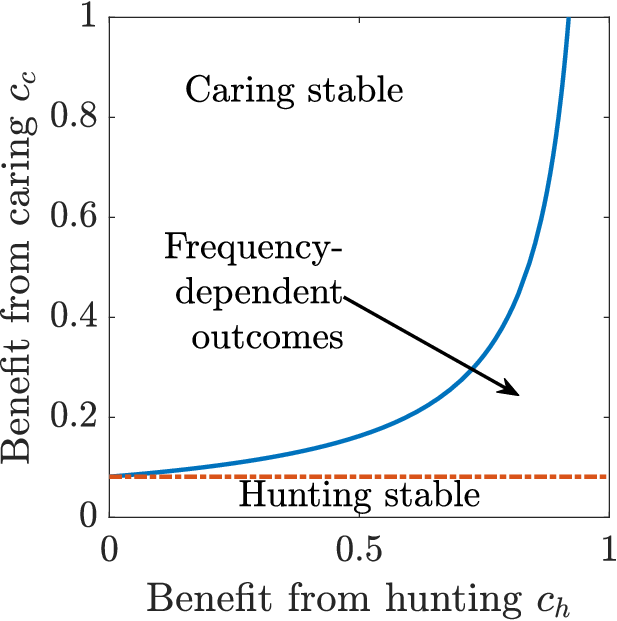}
    \caption{Competitive benefit $\alpha = 1.05$.}
    \label{subfig:lowalpha}
    \end{center}
\end{subfigure}
\hspace{20pt}
\begin{subfigure}[b]{0.4\textwidth}
    \begin{center}
    \includegraphics[width=\textwidth]{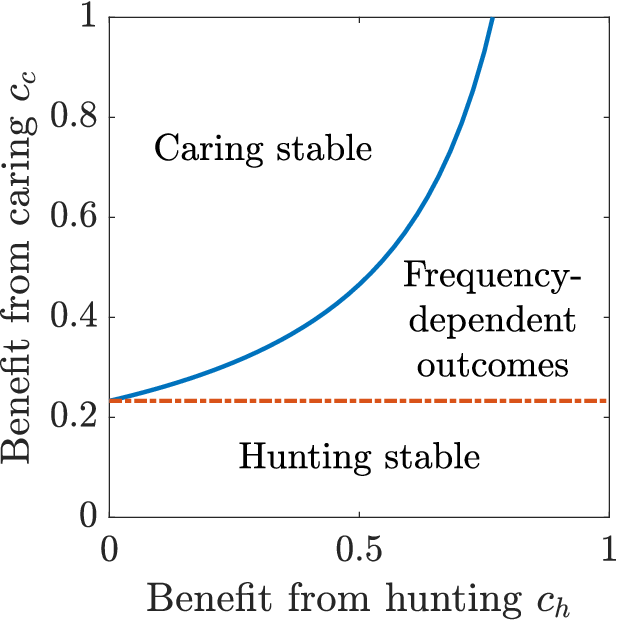}
    \caption{Competitive benefit $\alpha = 1.15$.}
    \label{subfig:highalpha}
    \end{center}
\end{subfigure}   
\end{center}
\internallinenumbers\caption{Equilibrium strategies at varying $c_c$ and $c_h$ for different values of competitive benefit, $\alpha$. The region above the solid blue curve corresponds to care overcoming large-game hunting, and the region below the red dash-dot curve to large-game hunting outcompeting care. Between the two curves is a region of frequency-dependent outcomes, where the initial population composition determines the outcome (as illustrated in Figure~\ref{fig:popsplit}). (\subref{subfig:lowalpha}) For very low competitive benefit,  $\alpha = 1.05$, the region where large-game hunting is successful is where the survival benefit of paternal care is low, but as the survival benefit of large-game hunting, $c_h$, increases, region of frequency dependence pushes into this region of care.  (\subref{subfig:highalpha}) For high competitive benefit, $\alpha = 1.15$, the region where only hunting is stable increases in size (the minimum value of benefit from paternal care, $c_c$, at which care can persist increases), and the minimum value of benefit from hunting, $c_h$, for which hunting can persist decreases, even for high $c_c$.}
  \label{fig:nonconstantbif1}
\end{figure}

\begin{figure}[ht!]
\begin{center}
\begin{subfigure}[b]{0.4\textwidth}
    \begin{center}
    \includegraphics[width=\textwidth]{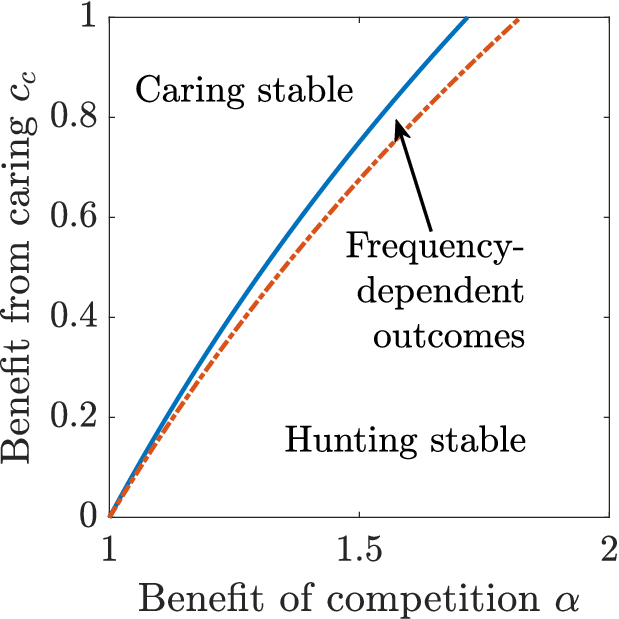}
    \caption{$c_h = 0.1$.}
    \label{subfig:lowch}
    \end{center}
\end{subfigure}
\hspace{20pt}
\begin{subfigure}[b]{0.4\textwidth}
    \begin{center}
    \includegraphics[width=\textwidth]{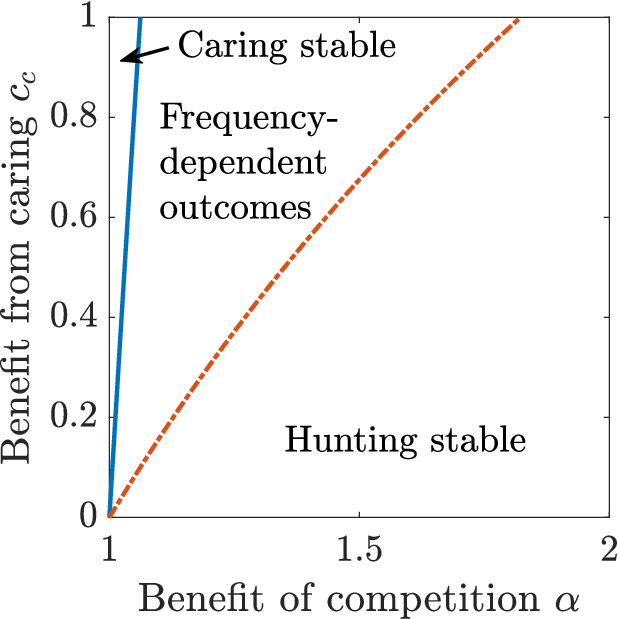}
    \caption{$c_h = 0.9$.}
    \label{subfig:highch}
    \end{center}
\end{subfigure}   
\end{center}
\internallinenumbers\caption{Bifurcation in $c_c$ and $\alpha$ for different values of benefit from large-game acquisition; (\subref{subfig:lowch}) low $c_h = 0.1$, and (\subref{subfig:highch}) high $c_h = 0.9$. Left of the solid blue curve is the region where care always outcompetes large-game hunting, and right of the red dash-dot curve is where hunting always outcompetes care; in between the two curves is the region of frequency dependence, where the initial population composition determines the outcome, as illustrated in Figure~\ref{fig:popsplit}. Figure~\ref{fig:bif1} illustrates the case when hunting confers no survival benefit to offspring, $c_h=0$.}
  \label{fig:nonconstantbif2}
\end{figure}

As some specific examples given certain realistic parameters, we further investigate the $c_h$-$c_c$ parameter space in Figure~\ref{fig:nonconstantbif1} for different values of $\alpha$ and the $\alpha$-$c_c$ parameter space in Figure~\ref{fig:nonconstantbif2} for different values of $c_h$, given equal initial proportions of carers and hunters. Note that Figure~\ref{fig:bif1} is similar to the plots in Figure~\ref{fig:nonconstantbif2} in the case $c_h=0$, causing the solid blue and the dashed red curves to coincide. In fact, the dashed red curve in both subfigures of Figure~\ref{fig:nonconstantbif2}, on the right of which competition is the only stable steady state, is fixed for all $c_h$ and has the equation $c_c = (\ln\alpha)/(\tau\mu)$. The solid blue curve in Figure~\ref{fig:nonconstantbif2} is given by $c_c = (\ln\alpha)/((1-c_h)\tau\mu)$, approaching $(\ln\alpha)/(\tau\mu)$ as $c_h\to0$. 

As the survival benefit from large-game hunting increases, i.e., $c_h$ increases, the direct care benefit $c_c$ required for paternal care to fully overcome large-game hunting (i.e., for the caring steady state to be stable) increases. In Figure~\ref{subfig:lowalpha}, where $\alpha = 1.05$, large-game hunting persists for low values of survival benefit to carers, below the red dash-dot line. As $\alpha$ increases, comparing Figure~\ref{subfig:highalpha} to \ref{subfig:lowalpha}, this region of large-game hunting persistence pushes further into the region of paternal care and frequency dependence. Equivalently, paternal care must deliver a substantially higher benefit in order to persist at all.

As the competitive benefit of large-game hunting increases, so does the frequency-dependent region. For $c_h \ge 1-(\ln\alpha)/(\tau\mu)$ (to the right of where the blue solid curve intersects $c_c=1$), there is always a set of initial conditions that allows large-game hunting to persist. This is the case regardless of $c_c$ -- even if carers provide the maximum amount of care, there remains some initial condition at which hunting can overcome care, though in this extreme case the benefit provided to all, $c_h$, must also be very high. 

Observe that the region of frequency dependence is given by $(\ln\alpha)/(\tau\mu) < c_c < (\ln\alpha)/((1-c_h)\tau\mu)$. Caring alone is stable for $c_c \ge (\ln\alpha)/((1-c_h)\tau\mu)$, and hunting alone is stable for $c_c<(\ln\alpha)/(\tau\mu)$. When life history scales with adult longevity $L$ (that is to say, age of sexual maturity $\tau\propto L$ and juvenile mortality $\mu\propto1/L$, for instance), the product $\tau\mu$ remains constant and the regions of stability remain fixed. Increasing juvenile mortality $\mu$ relative to age of sexual maturity $\tau$ reduces both threshold levels of $c_c$, but cannot eliminate the region of hunting-only stability (nor the region of frequency dependence). Decreasing juvenile mortality relative to the age of sexual maturity, then, \emph{increases} the region in which hunting alone is stable, and decreases the region in which caring alone is stable.

We then consider the behaviour of the bifurcation in the $\alpha$-$c_c$ plane (equivalent to the bifurcation shown in Figure~\ref{fig:bif1} from the simple model), given low ($c_h=0.1$ in Figure~\ref{subfig:lowch}) and high ($c_h=0.9$ in Figure~\ref{subfig:highch}) values of hunting survival benefit, $c_h$. These results are qualitatively similar to those of the analytic bifurcation for the constant offspring mortality case (Figure~\ref{fig:bif1}). As $c_h$ increases, the region to the left of the red dashed curve where paternal care can persist (provided certain initial conditions are satisfied) remains the same. However, the region to the left of the blue solid curve, where care is the only stable outcome, vastly reduces. In other words, the greater the benefit provided to all, the larger the region within which hunting can outcompete care. Despite the relaxation of the benefit to competing males, the competitive strategy is even more strongly selected for than in the general case.

\section{Discussion}

By modelling the payoffs to male reproductive effort in strategies of care and competition, we found regions within which either strategy outperformed the other. Even when paternal care prevented any offspring mortality (i.e., the offspring of carers necessarily survived), there remains scope for competition to outperform care and take over the population. Competitive strategies are robust to carers' benefits---selection acts more strongly in favour of competitive traits over those of caring. 

The payoff structure we employed modelled benefits as immediate marginal payoffs to surviving offspring (either better surviving offspring, or more to start with). Even with this simple structure, we found competition to outperform care for a wide range of parameters, and this was the case even in the absence of other effects that have been shown to increase selection for competition \citep[e.g. male-biased sex ratios in][]{Schacht16, Loo17a, Loo17b, Rose19}. The effects of sex ratio are not modelled here, but we note that human-like female scarcity that results from male-biased sex ratios \citep{Coxworth15} will only increase the selective pressure for competition, thus increasing the size of the region within which competition outcompetes paternal care. 

This persistence of competition at equilibrium even when the benefits to care are large emphasises how male-specific payoffs, that are shaped by anisogamy \citep{Lehtonen16}, can increase selection for competition over care. We model payoffs by assuming that males remain in the mating pool even while caring for offspring, and searching for other females. There is no pair bonding \citep[as modelled in][]{Schacht16,Loo17a,Loo17b} or alternative mechanism by which males are rendered unavailable, or unable, to mate. They are able to mate continually, a consequence of gametic differences. This aligns with other studies that make links between these payoffs and anisogamy \citep{Lehtonen16,Trivers72,Lehtonen14,Parker72}. Given the Fisher condition \citep{Fisher30,Queller97} that half the autosomal genes in diploids come from mother, half from father, males stand to profit from allocating effort into mating competition if the result is increased probability of fertilising a comparatively rare female egg \citep{Lehtonen16}. Those males who are outcompeted (in this case, carers) will fertilise fewer eggs, while competitive males secure more conceptions. This implies positive selection for male mating strategies, echoing the notion expressed in \citet[][p.88]{Darwin59} regarding the definition of ``sexual selection\dots depend[ing] not on a struggle for existence, but on a struggle between the males for possession of the females\dots the result [of which] is not death to the unsuccessful competitor, but few or no offspring". 

Our application of this simple payoff structure to the case of large-game hunting where resource benefits are provided to all was given in Section~\ref{sec:publicgood} and served as a case study demonstrating the utility of our first model. This shows that even when competition provides a benefit to all there is an increased selective pressure for that trait or behaviour.

Here, we model the potential survival benefit of the supply of the common good as well as increased offspring production to competers. Alternatively counting any action that affects offspring welfare as paternal care would only capture the survival benefit of large-game, a benefit that goes not only to the hunter's offspring but to the offspring of other males as well. This would ignore any other real payoffs that go to the individual hunter (such as increased offspring production). Our model captures both these payoffs to competition through large-game hunting. Our results also demonstrate the frequency dependence of these payoffs. When many hunters are present, the baseline offspring mortality rate is reduced (for all offspring regardless of strategy). Thus, the relative benefit to caring via provisioning of one's own offspring results in a smaller marginal benefit. The added public good payoff only \emph{increases} selection for the trait that increases offspring survival for all. This application not only demonstrates the model's versatility, but also provides some evidence of the show-off hypothesis \citep{Hawkes93, Hawkes02}. 

We identified a region of frequency-dependent stability in this model, and showed that large-game hunting may indeed outperform care even when the benefit to care is high. We further showed that as the effect of large-game hunting on the whole population increases (i.e., $c_h$ increases), the region within which hunting will necessarily succeed (where the hunting-only steady state is stable) increases. As the large-game meat acquired becomes increasingly more valuable to the welfare of offspring, care will diminish. 

Further, as a consequence of this frequency dependence, changes in environmental conditions and ecological conditions, which are not modelled here, may affect the survival benefit of either paternal care or large-game hunting changes or even the relative paternities gained through competition. These could then trigger a change in the preferred strategy across the entire population. This will indeed persist even if the conditions later return to their previous state. \citet{henshaw2019sex} demonstrated that sex roles can shift as ecological changes affect the costs and benefits of caring, though such transitions are likely to be rare. 

The simple model developed in this investigation sheds light on the importance of competition as a male reproductive strategy by characterising the marginal advantages of both paternal care and mating competition and investigating their interaction. When males are faced with the choice of investment in these alternative pathways, the model has shown that mating competition can win out for a wide range of parameters, even when the benefit of caring is high.

\section*{Acknowledgements}

The work of SLL was supported by the Australian Government Research Training Program Research Training Stipend. SLL, DR and PSK were supported by the Australian Research Council, Discovery Project (DP160101597).

\bibliographystyle{plainnat}
\bibliography{showbib}

\end{document}